\documentclass{aa}  

\usepackage{graphicx}
\usepackage{amssymb}
\usepackage{txfonts}
\usepackage{lipsum}
\usepackage{subcaption}         
\usepackage{lscape}             
\usepackage{placeins}           
\usepackage[colorlinks=true, citecolor=blue]{hyperref}
\hypersetup{
    citecolor  = blue,    
    linkcolor  = blue,   
    urlcolor   = blue,  
    pdfborder  = {0 0 0} 
}

\begin{document}

   \title{Constraining the outer boundary condition for the Babcock-Leighton dynamo models}

   \author{Yukun Luo\inst{1,2}
        \and Jie Jiang\inst{1,2}
        \and Binghang Li\inst{1,2}
        \and Zebin Zhang\inst{3,4}
        \and Ruihui Wang\inst{1,2}
        }

   \institute{School of Space and Earth Sciences, Beihang University, Beijing, People’s Republic of China\\
             \email{jiejiang@buaa.edu.cn}
            \and Key Laboratory of Space Environment Monitoring and Information Processing of MIIT, Beijing, People’s Republic of China\\
            \and Institute of Frontier and Interdisciplinary Science, Shandong University, Shandong, People’s Republic of China\\
            \and Institute of Space Sciences, Shandong University, Shandong, People’s Republic of China}

   \date{Received November 8, 2025}
 
  \abstract
   {The evolution of the Sun’s large-scale surface magnetic field is well captured by surface flux transport models, which can therefore provide a natural constraint on the outer boundary condition (BC) of Babcock–Leighton (BL) dynamo models.}
   {For the first time, we propose a zero radial diffusion BC for BL dynamo models, enabling their surface field evolution to align consistently with surface flux transport simulations.} 
   {We derive a zero radial diffusion BC from the Magnetohydrodynamic induction equation and evaluate its effects in comparison with two alternatives: (i) a radial outer BC, and (ii) a radial outer BC combined with strong near-surface radial pumping. The comparison is carried out both for the evolution of a single bipolar magnetic region and within a full BL dynamo model.}
   {The zero radial diffusion outer BC effectively suppresses radial diffusion across the surface, ensuring consistency between the evolution of the bipolar magnetic region in the BL dynamo and the surface flux transport model. With this outer BC, the full BL dynamo model successfully reproduces the fundamental properties of the solar cycle. In addition, the model naturally produces a surface magnetic field that is not purely radial, in closer agreement with solar observations.}
   {The physically motivated zero radial diffusion boundary condition paves the way for deeper insight into the solar and stellar cycles.}


   \maketitle

\section{Introduction} \label{sec:intro}
Boundary conditions (BCs) are essential for solving partial differential equations within a bounded domain. They ensure the mathematical well-posedness of the problem and guide the selection of physically relevant solutions. In the context of solar and stellar dynamo equations, BCs play a crucial role in shaping the behavior of large-scale magnetic fields. For instance, \citet{Schubert2001} and \cite{Jiang2006} suggested that imposing electrically conducting BCs at the core-shell interface facilitates the emergence of an oscillatory solution in $\alpha^2$ dynamo. \citet{Choudhuri1984} showed that changing the outer BC for the azimuthal field $B_\phi$ from $B_\phi=0$ to $\partial B_\phi/\partial r =0$ affects the cycle period of solar dynamo. \citet{Kapyla2010} found that for a given shear and rotation rate, the growth rate of the magnetic field is larger if a vertical field is adopted for the outer BC. Long-term observations and recent advances in understanding the solar surface magnetic field offer an opportunity to explore improvements to the outer BCs in solar dynamo models.  

Over the past decade, direct observations have increasingly supported the BL solar dynamo \citep{Babcock1961, Leighton1969}. The essence of this class of models is that the poloidal field is generated through the emergence of ARs and their subsequent evolution across the solar surface. A key feature of the BL solar dynamo is that it is observationally guided \citep{Cameron2023}. In the recently developed distributed-shear BL dynamo models, in which the toroidal field is generated in the bulk of the convection zone by the latitudinal shear \citep{Zhang2022}, the surface magnetic field plays a central role in modulating the cycle period and contributes to the time- and latitude-dependent regeneration of the toroidal field, thereby producing the butterfly diagram \citep{Jiang2025}. It is therefore natural to incorporate advances in our understanding of the solar surface magnetic field into the formulation of the outer BCs for the BL solar dynamo model.

The surface flux transport (SFT) models have demonstrated their glorious success in understanding the evolution of the solar surface magnetic field. The high-latitude radial component of the magnetic field $B_r$ is determined solely by processes that transport flux poleward across the surface from the activity belt at lower latitudes, behaving as passively advected corks. \cite{Leighton1964} draws the analogy between the motion of magnetic flux elements by the supergranulation over the surface and the motion of molecules in a gas, and gives the transport equation as 
\begin{equation}
	\label{eq:Leigton64}
	\frac{{\partial B_{r}}}{{\partial t}} = -\nabla  \cdot (\mathbf{u_h}{B_r})+ \eta_s \nabla^2_s B_r,
\end{equation}
where the horizontal field $\textbf{\textit{u}}_h$ corresponds to the differential rotation and the meridional flow at the solar surface, and $\nabla_s^2$ is the Laplacian operator on the surface of a sphere. \cite{Knobloch1981} and \cite{DeVore1984} demonstrated that the turbulent diffusivity, $\eta_s$, appears as a result of the ensemble averaging procedure, thereby providing a physical basis for its use in modeling the large-scale field. \cite{Wang1989} successfully applied the model to examine the polar field evolution of cycle 21. Since then, this model has been extensively employed to explore the temporal evolution of the solar surface magnetic field. Comprehensive discussions of the development and applications of this modeling framework are provided in the reviews by \cite{Jiang2014} and \cite{Yeates2023}. Most recently, \cite{Luo2025} demonstrated that the simulated spectra based on the model generally agree with observations for spherical harmonic degrees $l\lesssim60$.

When applying the SFT model to multiple solar cycles of varying strength, \cite{Schrijver2002} and \cite{Baumann2006} obtained a systematic drift of the polar fields and even the absence of polar field reversals. They argued that this disagreement with the observations reflects a conceptual deficiency in the original SFT model, and proposed that a radial diffusion term for $B_r$ should be included in Eq. (\ref{eq:Leigton64}) to remove the long-term memory of the surface flux. This raises the important question of whether radial magnetic flux indeed diffuse across the surface.

During the past decade, significant progress has been made in recognizing the cycle dependence of sunspot emergence. In particular, the cycle-averaged tilt angles is anti-correlated with the cycle strength \citep{Dasi2010, Jiao2021} and the cycle-averaged latitude is correlated with the cycle strength \citep{Li2003,Solanki2008,Jiang2011}. These dependencies act as effective nonlinear mechanisms for modulating the polar field generation and, consequently, the solar cycle, and are referred to as tilt quenching and latitude quenching \citep{Jiang2020,Petrovay2020,Karak2020}, respectively. By incorporating the cycle dependence of sunspot emergence into the source of the SFT model, \cite{Cameron2010} reproduced the polar field reversal over multiple solar cycles without the radial diffusion. Recently, \cite{Yeates2025} considered realistic AR configurations and successfully reproduced the evolution of polar fields across multiple solar cycles in line with the observations without the radial diffusion. Together, these studies suggest that additional radial diffusion across the surface is not required in the SFT process. Accordingly, for the BL-type dynamo models, the BCs should likewise be formulated to prevent the radial diffusion of the radial component of the magnetic field across the surface, ensuring consistency between the surface field evolution in BL dynamos and in SFT models. 

\cite{Cameron2012} made the first attempts to use the evolution of the SFT model to constrain near-surface properties of the BL dynamo. They suggest that a strong enough near-surface downward pumping with a vertical outer BC is required to reproduce surface field evolution from BL dynamo models in agreement with that from SFT models. Applying the constraint to BL dynamo models has lead to progress in understanding both solar and stellar cycle \cite[e.g.,][]{Jiang2013, Karak2016, Karak2017,Hazra2019,Zhang2022b}. Theoretically, near-surface radial pumping could arise from inhomogeneous turbulence or from a positive diffusivity gradient \citep{Zeldovich1957,Radler1968, Petrovay1994}. However, direct numerical simulations so far have primarily shown a downward transport of the large-scale magnetic field near the base of the convection zone \citep[e.g.,][]{Tobias1998, Ossendrijver2002,Kapyla2025}. Moreover, the combination of near-surface strong pumping and a vertical BC results in the presence of only the radial magnetic field component above 0.95R$\odot$ and in weak torsional oscillations \citep{Zhong2024}, both of which are inconsistent with observations. Furthermore, the study by \cite{Whitbread2019} compare the evolution of surface magnetic fields simulated from 3D BL dynamo and SFT models. Their results suggest that the disconnection between emerged active regions and the toroidal fields in the convection zone would be required for the consistency between dynamo models and the surface magnetic fields. The typical 2D BL dynamo naturally satisfies the requirement of disconnection, enabling us to focus on the role of the outer BC.

The objective of the paper is to propose a new constraint on the outer BCs for the BL dynamo, motivated by SFT modeling, which requires that no radial diffusion of the radial magnetic field component occurs across the solar surface. This constraint provides a physically consistent link between the SFT and BL dynamo frameworks and offers a step toward more realistic modeling of the solar cycle. 

The paper is organized as follows. In Sect. \ref{sec:derivation_BC}, we derive the zero radial diffusion BC for the BL dynamo. In Sect. \ref{sec:results}, we present the numerical results by applying the zero radial diffusion BC to the distributed-shear BL dynamo model. Finally, Sect. \ref{sec:conclusion} provides a summary of the results and a discussion of their implications.

\section{Derivation of the zero radial diffusion BC for the BL dynamo} \label{sec:derivation_BC}
The evolution of large-scale magnetic fields in the Sun is governed by the MHD induction equation, and the radial parts near the surface are expressed as
\begin{equation}\label{eq1}
	\frac{{\partial B_{r}}}{{\partial t}} = {\mathbf{e_r}} \cdot \nabla  \times (\mathbf{u} \times \mathbf{B} - \eta \nabla  \times \mathbf{B}),
\end{equation}
where $\textbf{\textit{u}}$ is the velocity of flow field and $\eta$ represents the turbulent diffusivity. We assume that $\eta$ varies only with radius and is independent of horizontal position.
Specifically, the first term of Eq. (\ref{eq1}) can be written as follows:
\begin{equation}\label{eq2}
	{\mathbf{e_r}} \cdot \nabla  \times \left( \mathbf{u} \times \mathbf{B}\right)  = \nabla  \cdot ({u_r}\mathbf{B_h}) - \nabla  \cdot (\mathbf{u_h}{B_r}),
\end{equation}
where the subscripts $r$ and $h$ correspond to the radial and horizontal direction, respectively. The term $\nabla\cdot(\textbf{\textit{u}}_h{B_r})$ represents the horizontal advection for $B_{r}$, and $\nabla\cdot({u_r}\textbf{\textit{B}}_h)$ is related to the emergence of sunspot, the source term in the SFT model \citep{Yeates2013,Jiang2014,Yeates2023}.

The last term of Eq. (\ref{eq1}) represents the diffusion process, and it can be expanded as follows:
\begin{equation}
	\label{eq3}	
	\begin{aligned}
		& - \mathbf{e_r} \cdot \nabla  \times (\eta \nabla  \times \mathbf{B}) =  \\  
		&\frac{\eta }{{{r^2}\sin \theta }}\frac{\partial }{{\partial \theta }}\left( \sin \theta \frac{{\partial {B_r}}}{{\partial \theta }}\right)  +     \frac{\eta }{{{r^2}{{\sin }^2}\theta }}\frac{{{\partial ^2}{B_r}}}{{\partial {\phi ^2}}}\\  
		&- \frac{\eta }{{{r^2}\sin \theta }}\frac{\partial }{{\partial \theta }}\left( \sin \theta {B_\theta } + r\sin \theta \frac{{\partial {B_\theta }}}{{\partial r}}\right)  \\  
		&- \frac{\eta }{{{r^2}\sin \theta }}\frac{\partial }{{\partial \phi }}\left( {B_\phi } + r\frac{{\partial {B_\phi }}}{{\partial r}}\right) .
	\end{aligned}
\end{equation}
The first two terms on the right-hand side of Eq. (\ref{eq3}) represent the horizontal diffusion, as shown in the 2D SFT model. Considering the axisymmetric assumption in our BL dynamo model, the last term involving the longitudinal gradient vanishes. The remaining third term represents an additional radial diffusion term \citep{Schrijver2002, Baumann2006}. This term is rewritten as
\begin{equation}\label{eq4}
	\begin{aligned}
		&- \frac{\eta }{{{r^2}\sin \theta }}\frac{\partial }{{\partial \theta }}\left( \sin \theta {B_\theta } + r\sin \theta \frac{{\partial {B_\theta }}}{{\partial r}}\right) \\ 
		& = - \frac{\eta }{{{r^2}\sin \theta }}\frac{\partial }{{\partial \theta }}\left(\frac{\partial }{{\partial r}}\left( r\sin \theta {B_\theta }\right)\right) \\ 
		& = - \frac{\eta }{{{r^2}\sin \theta }}\frac{\partial }{{\partial \theta }}\left(\frac{{{\partial ^2}}}{{\partial {r^2}}}\left( r\sin \theta A\right)\right) ,
	\end{aligned}
\end{equation}
where $A$ is magnetic vector potential in the $\phi$ direction, and the axisymmetric large-scale magnetic field $\textbf{\textit{B}}(r,\theta,t)$ is expressed in spherical coordinates as
\begin{equation}\label{eq: B_expasion}
	\textbf{\textit{B}}(r,\theta,t)=B_\phi(r,\theta,t) \mathbf{{e}_\phi}+
	\nabla\times
	\left[A(r,\theta,t){\mathbf{e_{\phi}}}\right].
\end{equation}

In the BL-type model, if we adopt the assumption that there is no radial diffusion across the solar surface to be consistent with the original SFT models satisfying Eq. (\ref{eq:Leigton64}), Eq. (\ref{eq4}) must vanish at the surface. This requirement corresponds to imposing the zero radial diffusion outer BC.  \cite{DeVore1984} derived Eq. (\ref{eq:Leigton64}) from the ideal MHD induction equation by decomposing the magnetic and velocity field into ensemble-averaged and fluctuating components, and by assuming that the large-scale magnetic field is predominantly radial so that horizontal field components are negligible ($B_\theta=0$ and $B_\phi=0$). In contrast, in our derivation of the zero radial diffusion outer BC, the radial field component naturally decouples from the horizontal components when the term $\nabla\cdot({u_r}\textbf{\textit{B}}_h)$ is neglected. This term is typically associated with flux emergence source in SFT models. As a consequence of this treatment, the latitudinal field component $B_\theta$ is permitted to exist at the solar surface, rather than being implicitly suppressed by the assumption of a purely radial field.

We note that Eq. (\ref{eq4}) equal to zero is a general form that makes no diffusion across the surface for poloidal fields, and a reduction form has been applied in dynamo models. \cite{Cameron2012} incorporate radial pumping in addition to the radial outer BC in the BL dynamo model as follows, 
\begin{equation}\label{eq5}
	{B_\theta } =-\frac{1}{r}\frac{\partial }{{\partial r}}(rA) = 0,r = {R_ \odot }.
\end{equation}
They obtain a surface evolution of the poloidal field that is consistent with that produced by SFT models. Radial pumping refers to the strong downward transport of magnetic flux near the solar surface, which tends to align the magnetic field predominantly in the radial direction. As a result, one obtains $B_\theta=0$ and $\frac{{\partial {B_\theta }}}{{\partial r}} = 0$ in the near-surface layer. The depth to which the relation $B_\theta=0$ holds depends on the penetration depth of the pumping. The near-surface pumping, along with the radial BC (Eq. (\ref{eq5})), makes the following equation approximately valid:
\begin{equation}\label{eq7}
	{\left. {\left({B_\theta } + r\frac{{\partial {B_\theta }}}{{\partial r}}\right)} \right|_{r = {R_ \odot }}} = 0.
\end{equation}
Then, Eq. (\ref{eq7}) at the surface is
\begin{equation}\label{eq8}
	{\left. \frac{\partial }{{\partial \theta }}\left[ \sin \theta\left( {B_\theta } + r \frac{{\partial {B_\theta }}}{{\partial r}}\right)\right] \right|_{r = {R_ \odot }}}=0.
\end{equation}
Therefore, the radial boundary condition combined with radial pumping can be regarded as a specific realization of the zero radial diffusion BC, obtained by setting both terms $B_\theta$ and $\frac{{\partial {B_\theta }}}{{\partial r}}$ equal to zero.

The general formulation of the zero radial diffusion boundary condition for the axisymmetric dynamo is:
\begin{equation}\label{eq9}
	{\left. {\frac{{{\partial ^2}}}{{\partial {r^2}}}\left( rA\right) } \right|_{r = {R_ \odot }}} = 0.
\end{equation}
Theoretically, the zero radial diffusion outer BC in an axisymmetric BL dynamo model could completely prevent magnetic diffusion across the surface, provided that the meridional flow does not subduct the poloidal field beneath the polar surface layers. Under this condition, the evolution of the surface radial field is expected to be consistent with that obtained from SFT models, after spatial averaging in the azimuthal direction. In the next section, we verify this expectation through numerical simulations.

\section{Numerical results} \label{sec:results}

This section aims to evaluate whether the new BC can reliably suppress the radial diffusion of the radial component of the magnetic field across the surface, thereby ensuring consistency between the surface field evolution in BL dynamos and that in SFT models. We first briefly overview the two-dimensional distributed-shear BL dynamo in Subsection \ref{subsec:BL-dynamo}. We next examine a simple case by comparing the time evolution of a single Bipolar Magnetic Region (BMR) under different setups of the outer BC in Subsection \ref{subsec:bmr}. We further examine the effects of different numerical implementations of the BC in Subsection \ref{subsec:decay}. The results of a full 2D distributed-shear BL dynamo model with the zero radial diffusion BC are presented in Subsection \ref{subsec:dynamo}.

\subsection{Overview of the 2D BL dynamo} \label{subsec:BL-dynamo}

The dynamo equations we solve for the toroidal field $B_\phi$ and the poloidal field represented by the magnetic vector potential $A$ are 
\begin{equation} \label{eq:dynamo_A}
	\frac{\partial A}{\partial t}+\frac{1}{s}\left(\mathbf{u_{p}}\cdot\nabla\right)(sA)
	=\eta\left(\nabla^{2}-\frac{1}{s^{2}}\right)A+S_{BL}, 
\end{equation}
\begin{equation}  \label{eq:dynamo_B}
	\begin{split}
		\frac{\partial B_\phi}{\partial t}+\frac{1}{r}\left[\frac{\partial(u_{r}rB_\phi)}
		{\partial r}+\frac{\partial(u_{\theta}B_\phi)}{\partial\theta}\right]= \\
		\eta\left(\nabla^{2}-\frac{1}
		{s^2}\right)B_\phi+ 
		s(\mathbf{B_{p}}\cdot\nabla\Omega)+\frac{1}{r}\frac{d\eta}
		{dr}\frac{\partial(rB_\phi)}{\partial r}, 
	\end{split}
\end{equation}
where the profiles of the meridional flow $\textbf{\textit{u}}_p$, angular velocity $\Omega$, turbulent diffusivity $\eta$, and BL-type source term $S_{BL}$ will be specified in the following subsections and $s=r \ sin\theta$. For the bottom boundary, we impose a perfect conductor condition, corresponding to $A = 0$, $\partial(rB_\phi) / \partial r = 0$ at $r = 0.65R_\odot$ \citep{Zhang2022}. 

For the outer BC, we consider three cases for comparison. Case 1 applies the newly proposed constraint for the vector potential $A$, Eq. (\ref{eq9}), with $B_\phi$=0 at $r=R_\odot$. Case 2 follows \cite{Cameron2012}, imposing a vertical magnetic field, i.e., $\partial (rA)/ \partial r = 0$, $B_\phi=0$ at $r = R_\odot$, along with strong near-surface radial pumping $\gamma$=25 ms$^{-1}$ as in their formulation. Case 3 adopts the same outer BC as Case 2 but omits the near-surface pumping, thereby permitting radial diffusion of the radial magnetic field across the surface. The potential-field outer BC is another commonly adopted form in dynamo modeling. However, since \citet{Cameron2012} demonstrated that it fails to reproduce results consistent with SFT modeling, we do not consider this case here.

The dynamo equations are numerically solved with the Crank-Nicolson scheme combined with an approximate factorization technique \citep{Houwen2001} developed at Beihang University. The code is second-order accurate in both space and time. For the default case, the computational grid is set to 129 points in both the radial and latitudinal directions. The scheme for dealing with the outer BC will be further addressed in the following subsections.

\subsection{Surface evolution of a single bipolar magnetic region in dynamo models} \label{subsec:bmr}

\begin{figure}[h!]
	\includegraphics[scale=0.35]{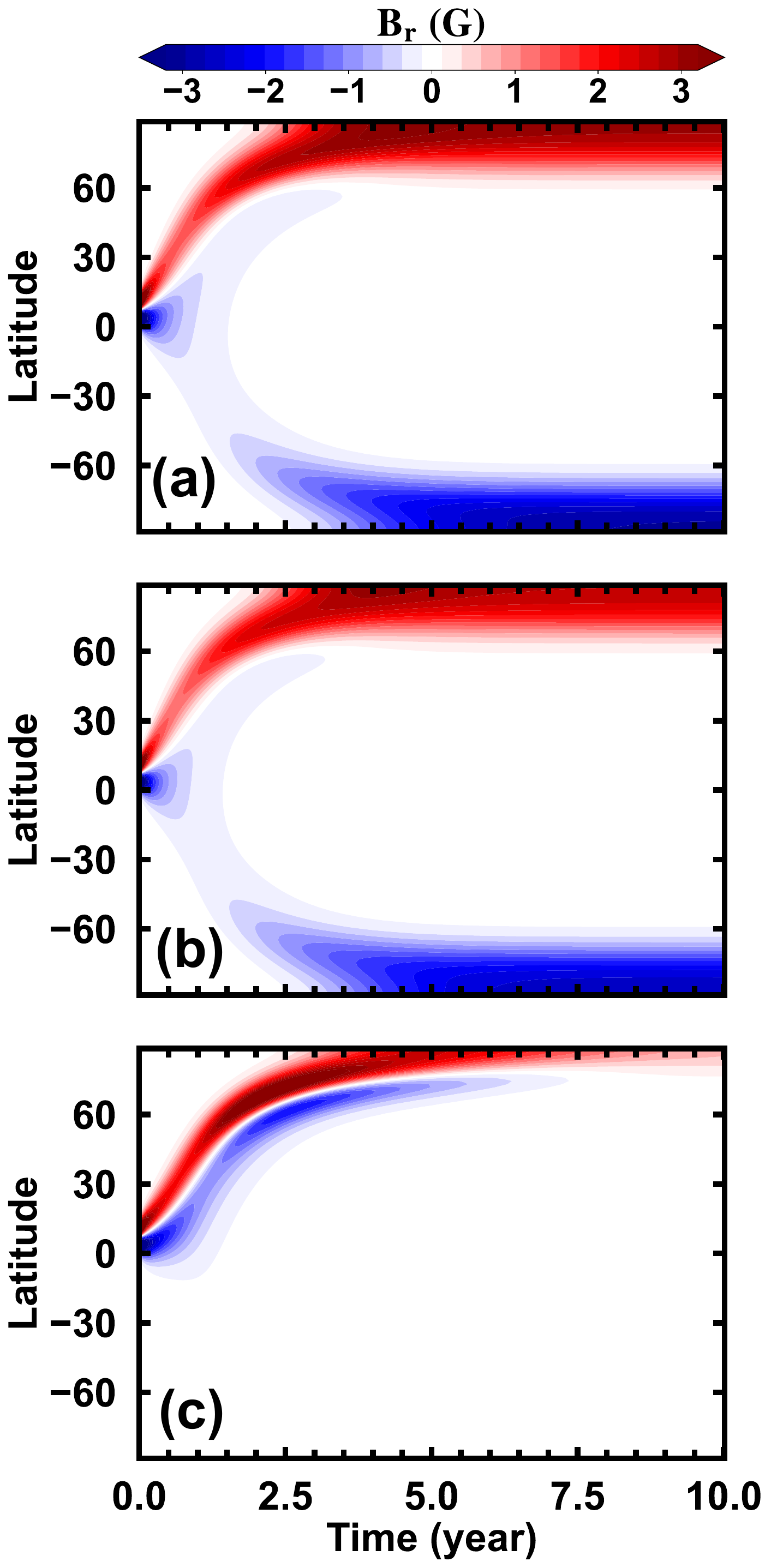}
	\caption{Temporal evolution of an initial BMR on the solar surface obtained from dynamo simulations under the three outer BC settings: (a) zero radial diffusion outer BC (Case 1); (b) radial outer BC along with strong near-surface radial pumping (Case 2); (c) radial boundary without near-surface radial pumping (Case 3).}
	\label{fig:bmr}
\end{figure}

\begin{figure}[htb!]
	\centering	
	\includegraphics[scale=0.10]{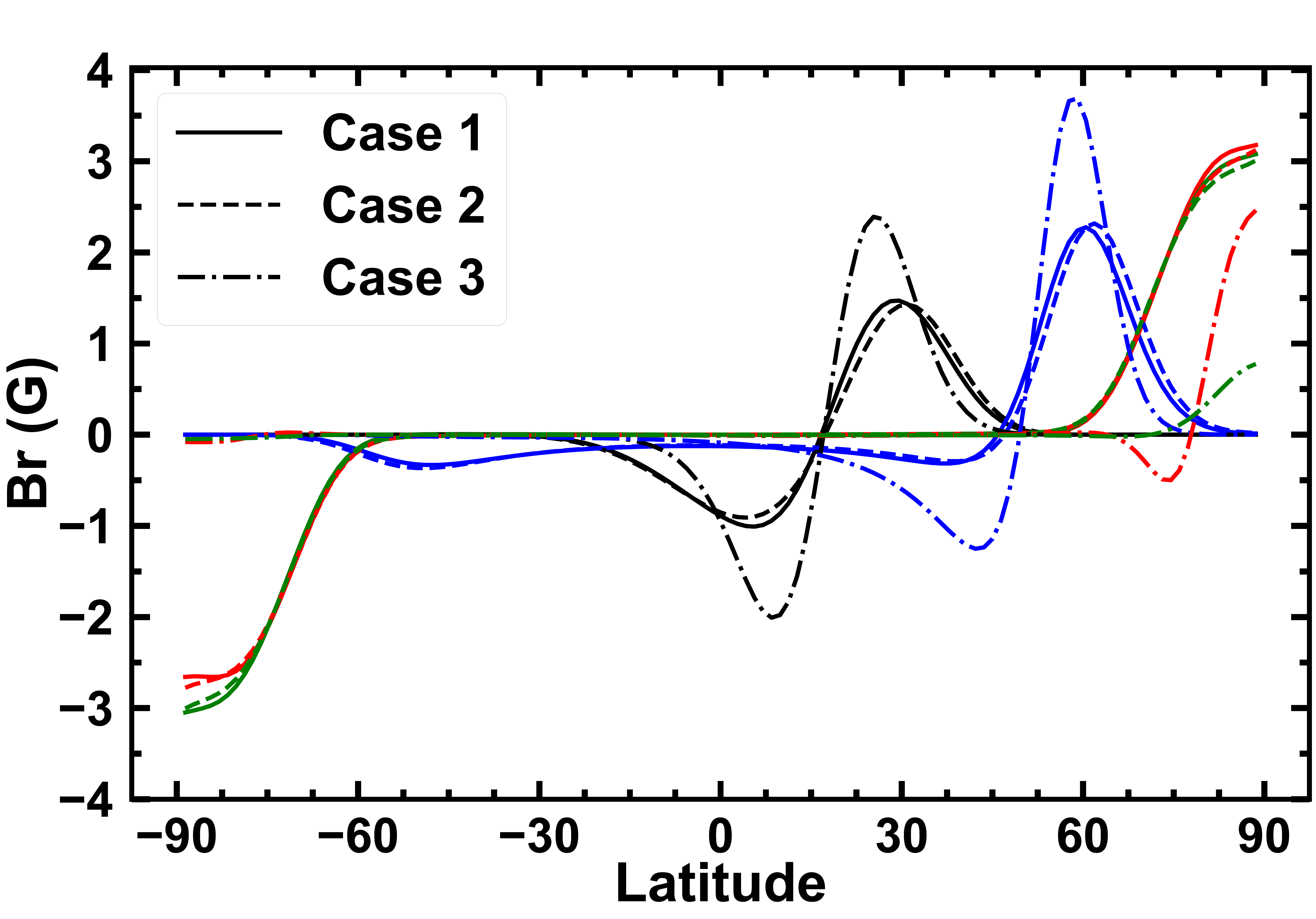}
	\caption{Comparison of the latitudinal distribution of the surface radial field $B_r$ from dynamo simulations with three different outer BC settings. Shown are snapshots at $t$ = 0.5 yr (black), $t$ = 1.5 yr (blue), $t$ = 6 yr (red), and $t$ = 10 yr (green). Solid, dashed, and dot–dashed curves correspond to Cases 1, 2, and 3 of the outer BC, respectively.}
	\label{fig:surface}
\end{figure}

In this section, we examine the evolution of a single BMR in the dynamo models under the three setups of the outer BC presented in the previous section. The initial field distributions of $A$ and $B_\phi$ follow Equations (9) and (10) of \cite{Cameron2012}, representing an isolated BMR centered at approximately 15$^\circ$ latitude. The meridional flow $\textbf{\textit{u}}_p$ and turbulent diffusivity $\eta$ profiles are the same as those in Section 2.1 of \cite{Cameron2012}, ensuring direct comparability. The angular velocity $\Omega$ and radial pumping profiles are set as those in \cite{Zhang2022}, which would not affect the results. The source term in the evolution equation of the poloidal field $S_{BL}$ is not considered, which allows us to assess whether the dynamo-based BMR evolution aligns with that from SFT models (cf. Figure 2 of \cite{Cameron2012}).

Within the framework of the SFT model (Fig. 2 of \cite{Cameron2012}), the evolution of the isolated BMR involves a small fraction of the leading-polarity flux (modulated by the tilt angle) being transported to the southern pole, while the equivalent amount of following-polarity flux is advected to the northern pole. In the absence of additional flux emergence and owing to the balance between the poleward meridional flow and the equatorward diffusion, the resulting polar field can remain stable for millennia. Accordingly, the temporal evolution of the BMR in dynamo simulations is expected to produce this behavior, provided that radial diffusion across the surface is effectively suppressed.

Figures \ref{fig:bmr} (a)-(c) compare the temporal evolution of the BMR in dynamo models under the outer BCs of Cases 1-3. Figures \ref{fig:bmr} (a)-(b), corresponding to Cases 1 and 2, present similar behaviors, consistent with that from the SFT model described above. In contrast, Case 3 (Fig. \ref{fig:bmr} (c)), which lacks near-surface pumping under the vertical BC, shows both polarities transported to the same pole. The ensuing cancellation between the two polarities leads to a rapid decay of the polar field. This implies a substantial reduction in dynamo efficiency and results in dynamo behavior that differs markedly from Cases 1 and 2. The potential-field outer boundary condition included in the dynamo produces a similar behavior to that of Case 3, as presented in \cite{Cameron2012}.

Figure \ref{fig:surface} presents a quantitative comparison of the latitudinal distribution of $B_r$ at four time snapshots, i.e., $t=0.5$ yr, 1.5 yr, 6.0 yr, and 10.0 yr. For all four times, Cases 1 and 2 exhibit nearly identical $B_r$ evolution, whereas Case 3 progressively diverges from their evolution with time. At $t=0.5$ yr, magnetic flux remains concentrated in both polarities for all cases. At $t=1.5$ yr, corresponding to the upper panel of Figure 3 in \cite{Cameron2012}, a small fraction of leading-polarity flux is transported to the southern hemisphere in Cases 1 and 2. At $t=6.0$ yr, corresponding to the lower panel of Figure 3 in \cite{Cameron2012}, equal-magnitude polar fields have been established for Cases 1 and 2, while flux of both polarities is concentrated within one hemisphere, resulting in weaker total unsigned flux due to cancellation. At $t=10.0$ yr, the polar fields remain nearly unchanged from $t=6.0$ yr, with northern and southern polar fields of 3.07 G and -3.04 G, respectively. In contrast, the magnetic flux in Case 3 is almost completely canceled. These results clearly imply that BL dynamo models employing typical outer BCs, e.g., vertical or potential-field BC, tend to produce decaying solutions.

In summary, the newly proposed zero radial diffusion BC and the vertical BC with strong near-surface pumping both suppress radial diffusion in BL dynamo simulations, which leads to the same time evolution of a single BMR. However, residual diffusion remains unavoidable due to numerical effects, especially a second-order derivative is involved in the zero radial diffusion BC. Hence, we examine how numerical schemes for implementing the second-order derivative BC affect the evolution of an initial global poloidal field in the next subsection. 

\subsection{Influence of numerical schemes on the evolution of an initial global poloidal field} \label{subsec:decay}

We use the magnetic field distribution at the end of the 10-year BMR evolution as the initial condition for the BL dynamo with the zero radial diffusion BC, which involves a second-order derivative. As in the previous subsection, no poloidal source term is included, and all other model parameters remain unchanged. In the absence of radial diffusion, the large-scale poloidal field is expected to remain stable for millennia because of the effect of the meridional flow. Near the equator, the poleward meridional flow inhibits cross-equatorial cancellation of the opposite radial flux, while in the polar regions, the balance between the poleward advection and the equatorward diffusion maintains the polar field \citep{DeVore1984}. To assess the influence of numerical schemes on suppressing radial diffusion, we apply the backward-difference scheme with varying orders of accuracy (second, third, and fourth) and two grid resolutions (129*129 and 259*259 in the radial and latitudinal directions). These tests enable us to quantify how discretization errors and grid resolution affect the long-term stability of the poloidal field.

Table \ref{table1} summarizes the ten-year change rates of the polar field strength at nearly 89$^\circ$ latitude for various grid resolutions and discretization orders of the second-order derivative BC. For a resolution of 129$\times$129 and second-order discretization accuracy, the radial magnetic field strength varies by approximately only $0.3\%$ after ten years, so it is sufficient to conclude that the assumption of negligible radial diffusion is satisfied. Increasing the discretization order of the BC from second to third and fourth significantly reduces the change rate by nearly an order of magnitude. Higher spatial resolution further mitigates the numerical errors. At a $259 \times 259$ resolution, the change rate decreases nearly an order of magnitude to 0.061\% with a second-order scheme, and to about 0.021\% with a fourth-order scheme. Furthermore, the quantitative rates of change depend on the surface diffusivity and meridional flow speed. These results demonstrate that higher-order discretization and finer grid resolutions are effective in improving the numerical accuracy of the second-order derivative BC, thereby suppressing the radial diffusion.  

Considering both accuracy and computational efficiency, we suggest that a $129 \times 129$ grid resolution combined with fourth-order discretization is an optimal numerical scheme for implementing the zero radial diffusion BC in dynamo simulations.

\begin{table}[ht]
	\centering
	\caption{Change rates of the polar field strength over 10 years for various grid resolutions and discretization accuracies of the second-order derivative BC.}
	\label{table1}
	\begin{tabular}{cccc}
		\hline \hline
		Grid resolutions & $O(\Delta x^2)$ & $O(\Delta x^3)$ & $O(\Delta x^4)$ \\ \hline
		129      & +0.3\%  & +0.039\%   & +0.034\%  \\
		259      & +0.061\%  & -0.015\%  & -0.021\% \\ \hline  
	\end{tabular}
\end{table}

\subsection{The distributed-shear BL Dynamo with the zero radial diffusion BC}\label{subsec:dynamo}

\citet{Zhang2022} developed the distributed-shear BL dynamo, which differs from the flux transport BL dynamo \citep{Choudhuri1995, Dikpati1999, Chatterjee2004}, as it does not rely on the subsurface meridional flow for flux transport \citep{Jiang2025}. The setup of the outer BC corresponds to Case 2 presented in Subsection \ref{subsec:bmr}, ensuring that the surface radial magnetic field evolves consistently with SFT models. Consequently, the poloidal field is dominated by large-scale components, primarily spherical harmonics $l$=1, 3, and 5 \citep{Luo2024}. In the bulk of the convection zone, the poloidal field is primarily latitudinal, enabling the generation of the toroidal field by the latitudinal shear. Hence, it is termed the distributed-shear model. Unlike flux transport dynamos, the distributed-shear model relies critically on the surface magnetic field. The surface polar field, shaped by the surface flux source and transport process, dominates the cycle period. Moreover, the evolving surface field is a factor contributing to the time- and latitude-dependent regeneration of the toroidal field that gives rise to the characteristic of the solar butterfly diagram. 

In subsect. \ref{subsec:bmr}, we have demonstrated that applying the zero radial diffusion BC to the BL model effectively captures the evolution of a BMR. In this subsection, we apply the same BC to the BL dynamo model of \cite{Zhang2022} to assess the resulting differences. Because we argue that the non-linear mechanisms in the BL dynamo primarily arise from tilt-angle quenching \citep{Dasi2010, Jiao2021} and latitudinal quenching \citep{Jiang2020, Petrovay2020, Karak2020}, we do not include the nonlinear algebraic quenching term in the source term $S_{BL}$ used by \cite{Zhang2022}. Instead, we investigate the critical solution of a linear dynamo model. The parameter $\alpha_0$ is set to its critical value $\alpha_c$, which is determined as the value at which the growth rate of the dynamo solution approaches zero. We adopt $\eta_{CZ}=2.32\times10^{11}$ cm$^2$s$^{-1}$ and keep other parameters identical to those in the reference model of \cite{Zhang2022}. The resulting critical value is $\alpha_c$=2.35 ms$^{-1}$.

\begin{figure*}[!htb]
	\centering	
	\includegraphics[width=17 cm]{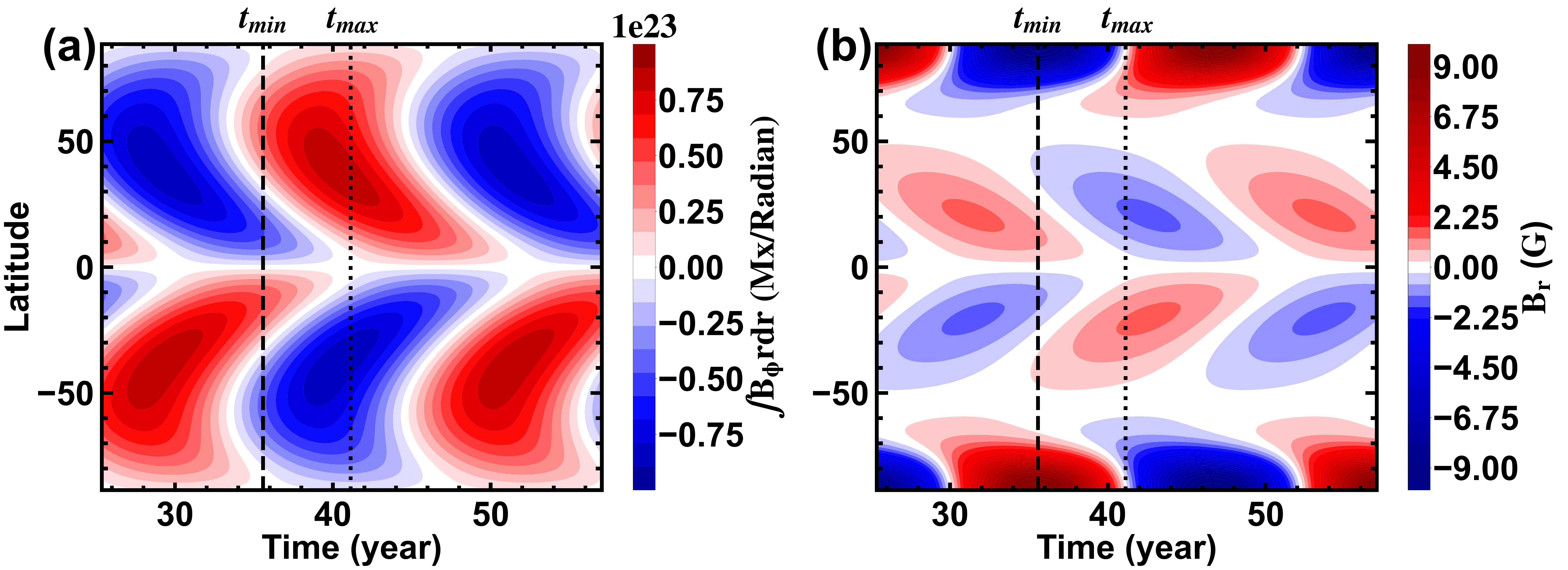}
	\caption{Dynamo solution with the zero radial diffusion outer boundary condition. (a) Temporal evolution of subsurface flux density of the toroidal field; (b) Temporal evolution of the surface radial fields. Cycle minimum ($t_{min}$) and cycle maximum ($t_{max}$) are indicated by the vertical dashed and dotted lines, respectively.}
	\label{fig:Bphi_r_t}
\end{figure*}

Figure \ref{fig:Bphi_r_t}a shows the time-latitude diagram of the subsurface toroidal flux density calculated by integrating the toroidal field over the range of 0.7- 1.0$R_\odot$ \citep[see][Eq. (8)]{Zhang2022}. Overall, the latitudinal migration pattern agrees with Fig. 4a of \cite{Zhang2022}, which corresponds to the outer BC of Case 2. In each cycle, the anti-symmetric toroidal field starts from about $\pm$55$^\circ$ latitudes along with a residual branch from the previous cycle near the equator. As the cycle progresses, the toroidal field presents both a poleward migration with a duration of about 11 years and an equatorward migration lasting about 18 years, consistent with the extended solar cycle \citep{Wilson1988,McIntosh2014}. The two-branch pattern differs from the results of FTD models, which typically produce concentrated flux above $\pm$70$^\circ$ latitudes due to strong radial shear in the polar tachocline. The poleward and equatorward branches of the toroidal field generated within the bulk of the convection zone in our model also provide a natural explanation for the observed torsional oscillation, driven by the Lorentz force associated with the cyclic magnetic field \citep{Zhong2024}.

A minor difference between Fig. \ref{fig:Bphi_r_t}a and Fig. 4a of \cite{Zhang2022} is that the maximum mean flux density in this result is approximately a factor of two lower. Nevertheless, this amplitude remains within the reasonable range estimated for the solar toroidal flux generation rate by \cite{Cameron2015}. The discrepancy originates from the distinct magnetic transport mechanisms near the surface. In \citet{Zhang2022}, strong downward pumping above 0.95$R_\odot$ rapidly transports the latitudinal field $B_\theta$ into deeper layers, where the latitudinal shear increases with depth (see Fig. 1d of \cite{Zhang2022}), thereby enhancing toroidal-field generation. In contrast, under the Case 1 outer BC adopted here, $B_\theta$ is advected downward gradually by the turbulent diffusion $\eta$, resulting in the weaker toroidal-field amplification. Consequently, the peak toroidal field strength in our model is about 150 G, compared to about 450 G under the Case 2 BC adopted by \cite{Zhang2022}. This value is also two orders of magnitude lower than the field strengths of buoyant magnetic loops in global convective-dynamo simulations \citep{Nelson2013,Nelson2014}. We emphasize that our result represents an azimuthally averaged toroidal field and does not account for the intermittent concentration of magnetic flux tubes as a result of the turbulent convection. 

Figure \ref{fig:Bphi_r_t}b shows the time-latitude diagram of the radial field $B_r$ at the solar surface. Overall, the result is consistent with Figure 4b of \cite{Zhang2022} and with the observed magnetic butterfly diagram \citep{Wang2025, Luo2025} in terms of the latitudinal migration pattern, the phase relation between the polar field and the activity cycle, and the polar field amplitude. Compared with observations, the absence of poleward surges in our model is likely due to the omission of ring doublets used to model individual active region emergence \citep{MunozJaramillo2010}. 

\begin{figure*}[htb!]
	\centering
	\includegraphics[width=17 cm]{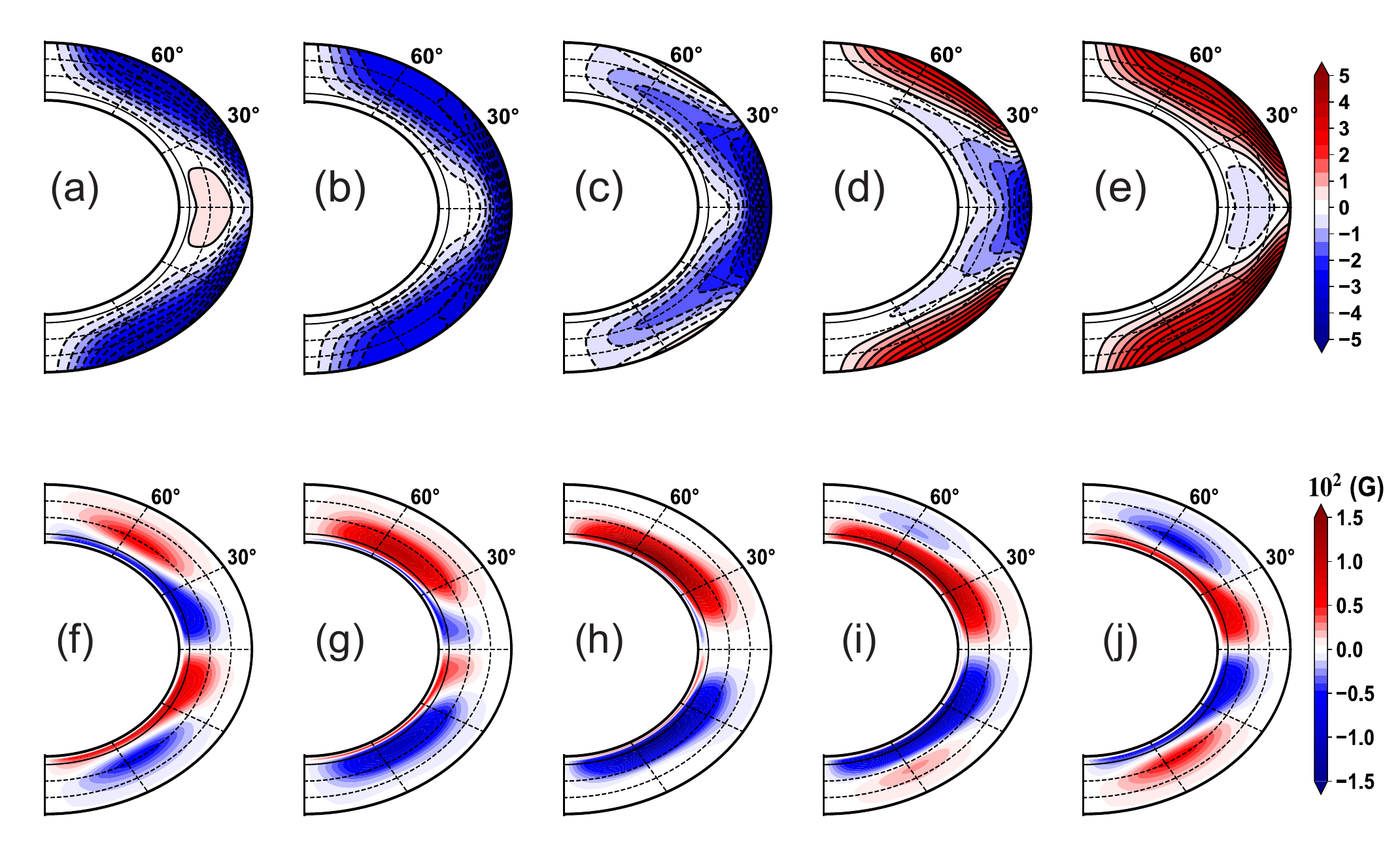}	
	\caption{Snapshots of the poloidal field (first row) and toroidal field (second row) over a dynamo cycle for the distributed-shear BL dynamo with the zero radial diffusion outer boundary condition. Solid (dashed) contours in the top panels correspond to clockwise (counterclockwise) poloidal field lines.}
	\label{fig:pol_tor_confi}
\end{figure*}

The top panels of Figure \ref{fig:pol_tor_confi} illustrate the evolution of the poloidal field over a simulated solar cycle. The surface magnetic field is not strictly radial: near the poles, the radial component dominates, whereas at lower latitudes the latitudinal component becomes significant. This behavior contrasts with the Case 2 outer BC adopted by \citet{Zhang2022}, where the surface field is enforced to be purely radial (see the top panel of their Fig. 6). Observations support that the polar field is nearly radial \citep{Sun2021}, and in the activity belts, the radial component typically dominates in regions of strong magnetic field, such as sunspots. Moreover, the field shown here is longitudinally averaged, corresponding to the magnetic butterfly diagram, where the mean radial field is only a few G. Meanwhile, a non-negligible large-scale latitudinal component $B_\theta$ is detected \citep{Solanki2006}. These observations explain the dominant latitudinal component at lower latitudes. Therefore, allowing a finite surface $B_\theta$ yields a more realistic representation of the large-scale solar surface field and can be regarded as an advantage of the current boundary condition relative to Case 2.

The low panels of Figure \ref{fig:pol_tor_confi} show the corresponding evolution of the toroidal field. The large-scale $B_\theta$ component generates a toroidal field within the bulk of the convection zone through the latitudinal differential rotation, consistent with the framework of the distributed-shear BL dynamo. For each cycle, the toroidal field originates near $\pm$55$^\circ$ latitudes, and subsequently migrates both equatorward and poleward, driven by the latitudinal dependence of the latitudinal differential rotation. This behavior contrasts with that of FTD models. In addition to the strength of the poloidal field, turbulent diffusion plays a key role in the cyclic regeneration of the toroidal field by enabling flux cancellation between successive cycles for the rising phase and across the equator for the decline phase \citep{Cameron2016}. These characteristics are generally consistent with those obtained using the Case 2 outer BC, except that the toroidal field here is slightly weaker, owing to the reduced strength of the corresponding poloidal field.

\section{Discussion and Conclusions} \label{sec:conclusion}
In this paper, we have introduced a new outer boundary condition, a zero radial diffusion boundary condition, for the BL dynamo to ensure that the surface field evolution is consistent with observations. We have verified the effectiveness of this boundary treatment using both an individual BMR case and a full dynamo simulation. The BC is approximately equivalent to adopting a vertical-field condition combined with near-surface radial pumping, as proposed by \cite{Cameron2012}, but offers two key advantages. First, the need to introduce radial pumping as an additional physical process is eliminated, in line with the spirit of Occam’s razor. Second, it naturally produces a surface magnetic field that is not purely radial, which is more consistent with solar observations. Our preliminary analysis further indicates that the new outer BC enables a more realistic reproduction of the surface distribution and temporal evolution of torsional oscillations, outperforming the Case 2 outer BC evaluated in \cite{Zhong2024}.   

With the surface field correctly maintained, the large-scale poloidal field dominates near the surface, leading to the generation of toroidal field primarily in the bulk of the convection zone through latitudinal differential rotation, which is in the same scenario of distributed-shear BL dynamo \citep{Zhang2022, Jiang2025} and the original BL dynamo \citep{Babcock1961, Leighton1969}. The realistically reproduced surface magnetic field also significantly enhances the dynamo efficiency by suppressing radial diffusion, in contrast to the widely used vertical-field outer BC and the potential-field matching BC. Our results highlight that the surface magnetic field produced by the emergence and evolution of active regions is a fundamental element of the BL dynamo loop. Yet, most previous BL dynamo models, such as the FTD models, did not incorporate realistic constraints on the surface boundary, resulting in surface field evolution inconsistent with observations and, consequently, altered dynamo behavior in the convection zone.

The outer BC matching a potential field outside has been widely adopted in previous dynamo models. However, it is well recognized that the solar photosphere and corona exhibit non-potential magnetic structures \citep[e.g.,][]{Gary1987, Seehafer1990, Pevtsov1997}. To quantify the non-potential and helical properties of the coronal magnetic field, \cite{Pipin2025} recently considered the harmonic magnetic field as the outer BC, originally proposed by \cite{Bonanno2016}, into dynamo simulations. They argue that the non-potential and helical nature of the coronal magnetic field originates directly from the dynamo region \citep[e.g.,][]{Kapyla2010,Warnecke2011}. Similarly, weak horizontal surface fields are also permitted in the model of \cite{vanBallegooijen2007}, who reproduce the observed spreading of active region magnetic flux across the solar surface by coupling magnetic field evolution in the convection zone with that in the corona. An alternative interpretation attributes these non-potential and helical properties to surface processes, including differential rotation, meridional flow, supergranular diffusion, and the emergence of twisted active regions \citep[e.g.,][]{Yeates2014,Yeates2022}. The hemispheric pattern of filaments \citep{vanBallegooijen1998,Yeates2007} has been successfully reproduced by coupling SFT models to coronal evolution models, in which the photospheric and coronal fields co-evolve self‐consistently. Our axisymmetric BL dynamo simulations reproduce the behavior of reduced one‐dimensional SFT models, where the large-scale radial field is spatially averaged in the azimuthal direction. We therefore anticipate that, when the zero radial‐diffusion outer BC is implemented in a three‐dimensional kinematic distributed‐shear BL dynamo, its surface magnetic evolution will be consistent with full SFT simulations. Consequently, the non-potential and helical coronal magnetic structures may also be reproduced in future extensions of our work.

In addition, we note that the zero radial diffusion BC we introduce involves a second-order radial derivative. The Cauchy problem for the second-order partial differential equations in Eqs. (\ref{eq:dynamo_A})–(\ref{eq:dynamo_B}) is typically posed with Dirichlet, Neumann, or Robin boundary conditions. There is no general existence theorem guaranteeing a well-posed solution when a second-derivative Neumann-type boundary condition is imposed. Nevertheless, higher-order boundary conditions, including second-order ones, have been widely used in practice when physically motivated. A common example arises in wave‐equation problems, where such conditions have been successfully implemented \citep{Engquist1977,Bamberger1990,Poinsot1992}. These physically grounded boundary treatments have demonstrated both necessity and robustness in capturing the essential dynamics of the system.

\begin{acknowledgements}
We thank the anonymous referee for the valuable comments and suggestions, which helped us to improve our manuscript. The research is supported by the National Natural Science Foundation of China (grant Nos. 12425305, 12350004, and 12173005).
\end{acknowledgements}

\bibliographystyle{aa.bst}
\bibliography{references}

\end{document}